\documentclass[letterpaper,twocolumn,10pt]{article}
\usepackage{usenix2019_v3}
\usepackage{comment}

\usepackage{amsmath,amsfonts,amsthm} 
\usepackage{array}
\usepackage{tabularx}
\usepackage{graphicx}
\graphicspath{{./}}
\usepackage{xspace}
\usepackage{color}
\usepackage{listings}
\usepackage{url}
\usepackage{courier}
\usepackage{enumerate}

\usepackage{lmodern}
\usepackage[T1]{fontenc}
\usepackage{geometry}
\usepackage{csquotes}

\newcommand{\optane}{Optane-PMM\xspace}
\newcommand{\pmem}{PM\xspace}

\newcommand{\code}[1]{\begin{ttcodefont}#1\end{ttcodefont}}
\newcommand{\smallcode}[1]{\begin{ttsmallcodefont}#1\end{ttsmallcodefont}}

\begin{document}


\title{\Large \bf A High-Performance Persistent Memory Key-Value Store \\with Near-Memory Compute}

\author{
  {\rm Daniel G. Waddington}\\
  daniel.waddington@ibm.com \\
  IBM Research Almaden
\and
  {\rm Clem Dickey}\\
  dickeycl@us.ibm.com \\
  IBM Research Almaden
\and
  {\rm Luna Xu}\\
  xuluna@ibm.com \\
  IBM Research Almaden 
\and
  {\rm Moshik Hershcovitch}\\
  moshikh@il.ibm.com \\
  IBM Research Almaden
\and
  {\rm Sangeetha Seshadri}\\
  seshadrs@us.ibm.com \\
  IBM Research Almaden
} 

\maketitle

\begin{abstract}

MCAS (Memory Centric Active Storage) is a persistent memory tier for
high-performance durable data storage. It is designed from the
ground-up to provide a key-value capability with low-latency
guarantees and data durability through memory persistence and
replication.

To reduce data movement and make further gains in performance, we
provide support for user-defined ``push-down'' operations (known as
Active Data Objects) that can execute directly and safely on the
value-memory associated with one or more keys. The ADO mechanism
allows complex pointer-based dynamic data structures (e.g., trees) to
be stored and operated on in persistent memory.

To this end, we examine a real-world use case for MCAS-ADO in the
handling of enterprise storage system metadata for Continuous Data
Protection (CDP). This requires continuously updated complex metadata
that must be kept consistent and durable.

In this paper, we i.) present the MCAS-ADO system architecture, ii.) show
how the CDP use case is implemented, and finally iii.)
give an evaluation of system performance in the context of this
use case.
\end{abstract}

\section{Introduction}
While the notion of persistent memory has been around for many
decades, its widespread availability had not arisen until recently.
By 2021, Intel's Optane 3D-Xpoint Persistent Memory (PM) DIMM market is
projected to reach \$1B in revenue~\cite{mkwv2020}.  Its value
proposition, compared to DRAM, is centered around lower-cost per GB,
larger capacity, lower-energy consumption through refresh
elimination, and data retention (persistence) in power-failure and
reset events.

\pmem is \textit{byte-addressable} in that it is directly accessed via
load-store instructions of the CPU. The current generation
of Intel's Optane PM operates at a read-write latency of around
\(300ns\)~\cite{spectra2020, izraelevitz2019basic}. Even though this
is slower than DRAM access latencies (\(\sim100ns\)) it is at
least 30x faster than state-of-the-art storage (e.g., NVMe SSD). Capacity of
\pmem is also about 8x that of DRAM\footnote{At least for 3D-XPoint, which
 is based on lattice-arranged Phase Change Memory (PCM).}.


As with storage, in order to enable data sharing and availability \pmem
must also be provisioned as a network-attached resource.  Today, a
growing number of key-value stores use memory for performance and
replication for
reliability~\cite{10.1145/3030207.3053671,Stonebraker2013TheVM,10.14778/1454159.1454211,10.1145/2806887,10.5555/2616448.2616488,10.1145/2872887.2750416,10.1145/2619239.2626299,10.5555/2616448.2616486}.
More recently, key-value stores that support \pmem are emerging in
academia\cite{227826,Tsai2020DisaggregatingPM,choi2020observations}.
However, this existing work is limited in its ability to take full
advantage of near-data compute.


The key contributions of the paper are as follows:
\begin{enumerate}[i.]
\item Extensions of the ``plain-old'' network-attached MCKVS key-value
  store~\cite{waddington2020ispass} that provide flexible Active
  Data Object (ADO) capabilities allowing arbitrary compute to be safely
  \textit{pushed down} into the storage system.
\item Demonstration of the solution in a real-world use case based on
  a fast and durable indexing data structure for Continuous Data
  Protection (CDP) in enterprise storage systems.
\item Evaluation of general throughput and latency performance/scaling.
\item Comparison of performance for the ADO paradigm versus a plain-old
  key-value paradigm indicating a 43\% performance improvement in
  our given use case.
\end{enumerate}
  
The rest of the paper is organized as follows:
Section~\ref{sec:design} provides a background on MCAS, which is
derived from MCKVS~\cite{waddington2020ispass}, and then describes the
architecture of Active Data Objects (ADO)
extension. Section~\ref{sec:usingmcas} details the application of ADO
extensions to various functions within an over-arching CDP index
service.  Section~\ref{sec:eval} presents the evaluation
results. Section~\ref{sec:related} and ~\ref{sec:con} summarize the
related work and conclusions respectively.

\section{MCAS Design}
\label{sec:design}

At a fundamental level, MCAS is a network-attached key-value store
that is enhanced to allow flexible in-store operations directly on persistent
memory. It is a \textit{sharded} architecture
enabling a lock-free design with the restriction of only allowing
$1:N$ shard-to-pool mapping (i.e. any pool can only be serviced by a
single shard). Each shard is associated with a set of persistent
memory regions and CPU resources.  Memory regions can be either a
device DAX (devdax) partition or file-system DAX (fsdax)
file\footnote{Use of file-system DAX requires On-Demand Paging
  hardware in the RDMA NIC.}.  At the socket level, \optane is
configured in App-Direct mode and data is striped across six DIMMs in
order to achieve maximum socket performance.

\begin{figure}
\centering
\includegraphics[width=0.8\linewidth]{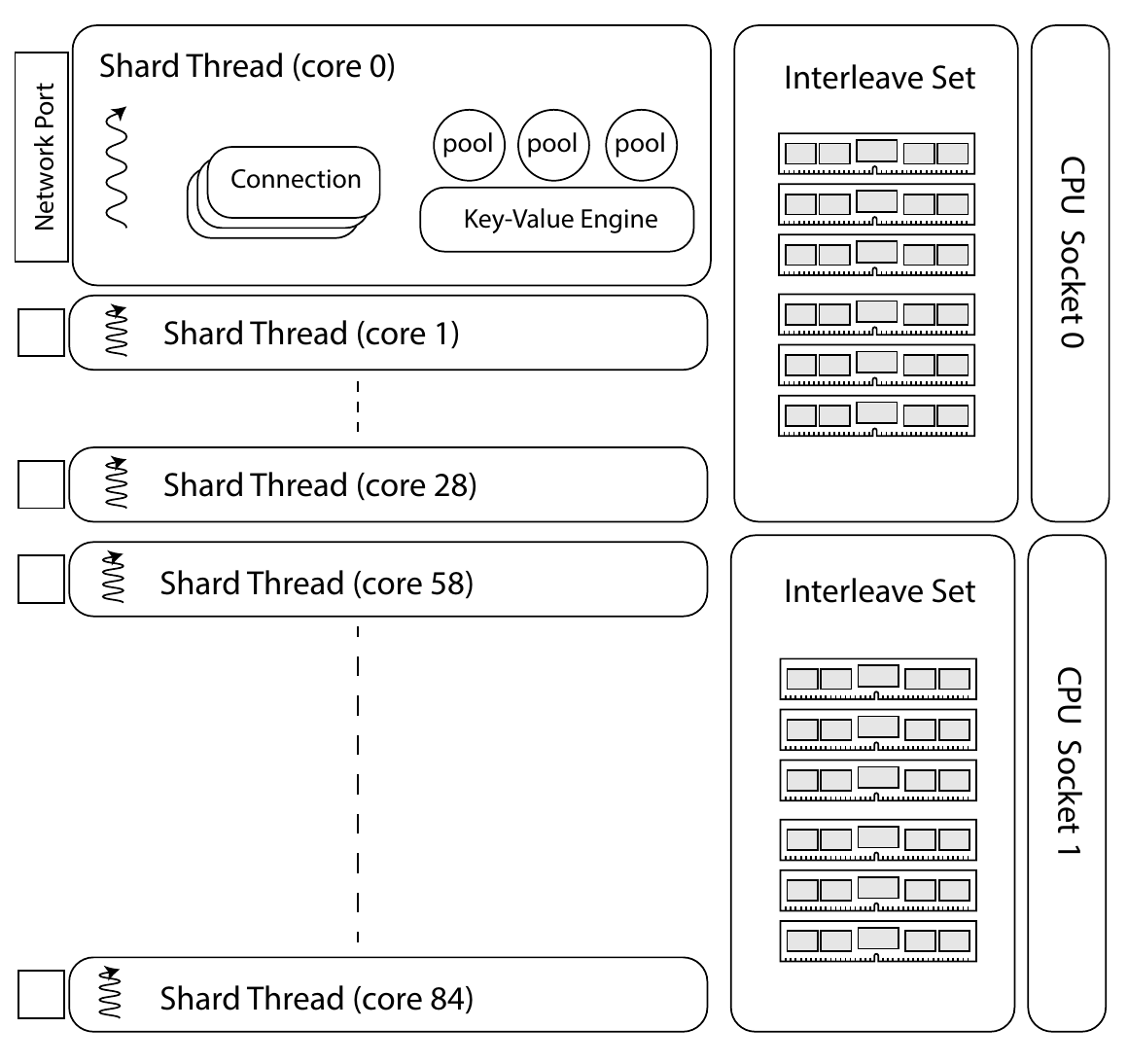}
\caption{MCAS Sharding Architecture}
\label{fig:shardarch}
\end{figure}

Each shard also maintains a single network connection end-point (port).
MCAS uses \textit{libfabric} with the RDMA verbs or TCP/IP sockets
provider. For the work in this paper, the RDMA provider is used since this
offers high-performance kernel-bypass access to the network.

Data is arranged in \textit{pools} that represent a set of persistent
memory resources, which are allocated from a coarse-grained allocator
managing the shard's memory regions. Pool memory is used for the hash
table index, keys and values, and can be dynamically expanded. Each
pool represents a logical set of key-value pairs in a single key
namespace.


MCAS supports variable key and value sizes. This
means that a heap-allocator is necessary to support variable-length
region allocation. To support high-rates of key-value pair insertion
and deletion, MCAS maintains a heap allocator for key-value
data in volatile (DRAM) memory. 

The core of the persistent key-value index is a hopscotch hash
table~\cite{HerlihyST08}, which maps keys to values. The hash table for
a given key-value pool is maintained in persistent memory.

Further detail on MCAS core (MCKVS) is given
in~\cite{waddington2020ispass}.

\subsection{Active Data Objects}
\label{sec:eval}

A key feature of MCAS is its ability to support in-store operations
via Active Data Object (ADO) plugins. This allows the advanced user
to associate in-store compute with a given pool.

ADOs are ``sandboxed'' compute that allow for generalized in-place
execution on persistent memory that is managed by MCAS.  They execute
outside of the shard process, which maintains the key-value index, so
that their scope of data access can be limited. ADOs enable the layering of
both domain-specific (e.g., matrix multiply) and generalized services
(e.g., replication, versioning) over the basic key-value store.

ADOs are implemented as plugins.  The plugins are \textit{invoked} via
the following client-side operation:

\begin{lstlisting}
status_t invoke_ado(
 const IMCAS::pool_t pool,
 const std::string& key,
 const void* request,
 const size_t request_len,
 const ado_flags_t flags,
 std::vector<ADO_response>& out_response,
 const size_t value_size = 0);
\end{lstlisting}

On instantiation, the shard process maps the regions of persistent
memory corresponding to the pool into the ADO process. Thus, the ADO
process has complete visibility of the pool memory (the security
boundary is the pool).

The invoke operations incorporate an \textit{opaque request}.  This
passes from the client, through the shard process and then into the
ADO process via a user-level IPC connection. Ultimately, on receiving
a request, the ADO process invokes the plugin's \code{do\_work}
operation:

\begin{minipage}{\linewidth}
\begin{lstlisting}[language=C++]
status_t do_work(
 const uint64_t work_id,
 const char* key,
 const size_t key_len,
 IADO_plugin::value_space_t& values,
 const void* in_work_request,
 const size_t in_work_request_len,
 const bool new_root,
 response_buffer_vector_t& response_buffers);
\end{lstlisting}
\end{minipage}

The \code{key} and \code{key\_len} parameters are provided as a
reference point because actions are normally logically associated with a given
key-value pair. Nevertheless, the ADO plugin has full visibility of
the key-value space within the whole pool. In order to interact with the
memory and data therein, the ADO plugin uses call-backs to the
shard process to manage memory, see Table~\ref{tab:callbacks}.

\begin{table}[ht]
\begin{centering}
\begin{tabularx}{1.0\linewidth}{
>{\setlength{\hsize}{.4\hsize}\raggedright\footnotesize}X
>{\setlength{\hsize}{.6\hsize}\raggedright\arraybackslash\footnotesize}X }
 \hline
 \textbf{Functions} & \textbf{Description} \\
 \hline
 \smallcode{create/open/erase} & Key-value management \\
 \smallcode{resize value} & Resize existing value \\
 \smallcode{allocate/free memory} & Pool memory management \\
 \smallcode{get ref vector} & Get vector of key-value pairs \\
 \smallcode{iterate} & Iterate key-value pairs \\
 \smallcode{find key} & Scan for key through secondary index \\
 \smallcode{get pool info} & Retrieve memory utilization etc. \\
 \smallcode{unlock} & Explicitly unlock key-value pair \\
 \hline
\end{tabularx}
\caption{ADO plugin callback API}
\label{tab:callbacks}
\end{centering}
\end{table}

ADO plugins can also be layered. Signals and data can be exchanged
between layers via the response buffer vector and/or shared memory of
the pool (e.g., a special key-value pair). Layered plugins are
invoked in a round-robin manner.

\begin{figure}
\centering
\includegraphics[width=0.8\linewidth]{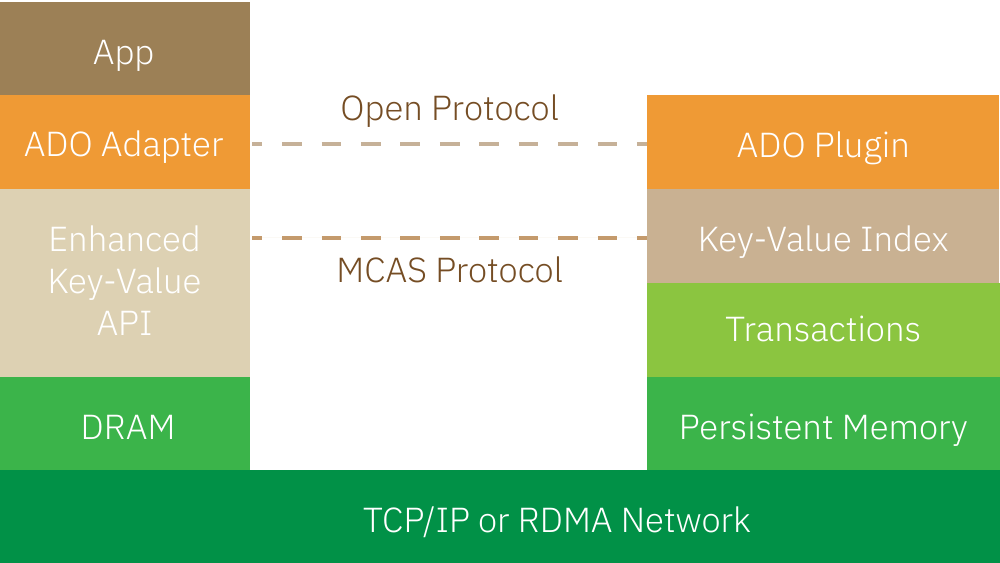}
\caption{MCAS Protocol Stack}
\label{fig:protostack}
\end{figure}

A subtle design characteristic of the ADO plugin architecture, is that
only the plugin itself can interpret the opaque message. Thus, the ADO
plugins and client-side libraries (that construct the requests) are
packaged as part of a readily extensible \textit{open protocol} (see
Figure~\ref{fig:protostack}. The advantage of this design is that any
arbitrary functionality can be implemented in the ADO plugin, e.g.,
replication, encryption, data structure manipulation, tiering. The
opaque request can be program (e.g., compiled or interpretable code)
or data (e.g., encrypted message).

MCAS also supports an \code{invoke\_put\_ado} operation,
which allows the client to ``put'' a value into the store prior to
invoking the ADO plugin.

\begin{figure}
\centering
\includegraphics[width=1.0\linewidth]{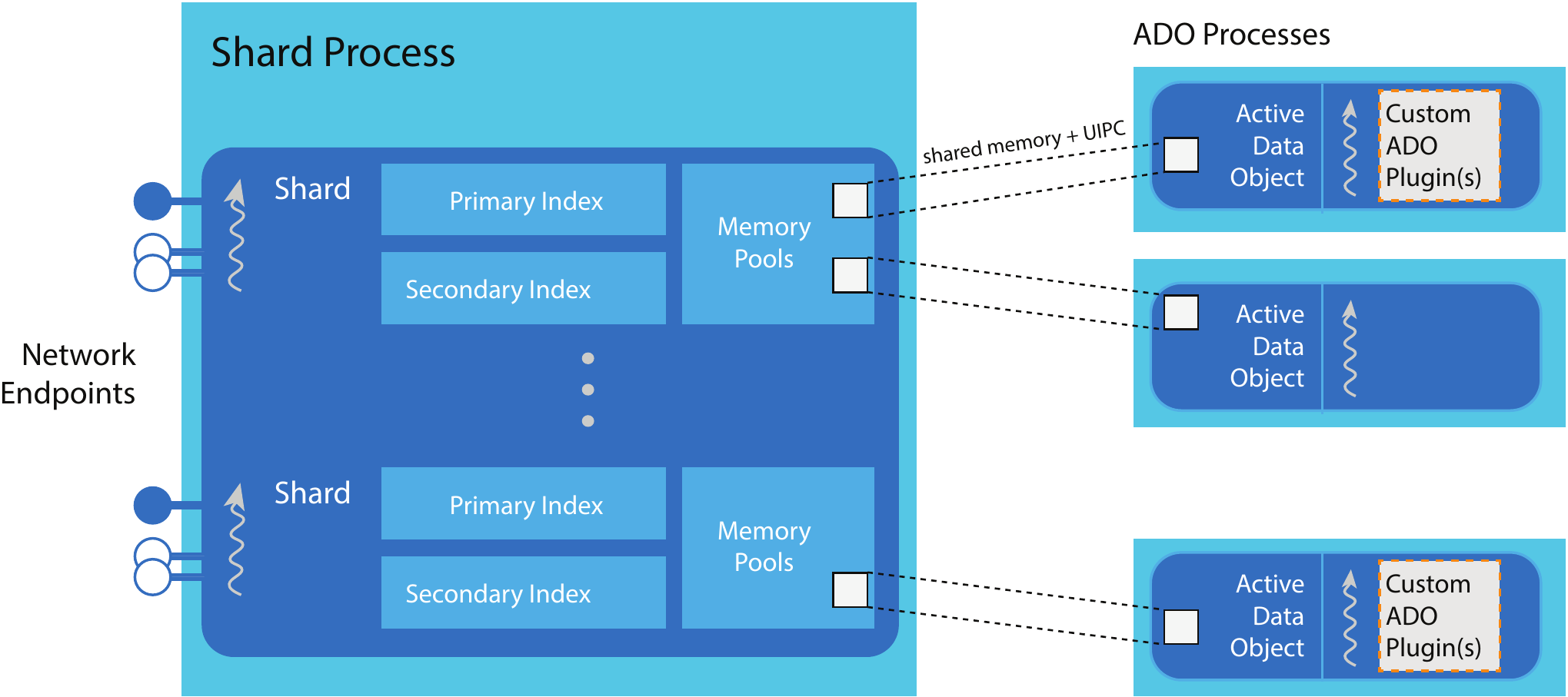}
\caption{ADO Architecture}
\label{fig:adoarch}
\end{figure}


Any data that is operated on by the ADO plugin must be
\textit{crash-consistent}~\cite{10.1145/2800695.2801719}. This means
that if the system fails at any point of time, the data structure can
at least be recovered to a \textit{transactionally complete} state.
The most common approach to realizing crash-consistency is to use an
\textit{undo-log} or \textit{redo-log}.  

Other more advanced techniques that rely on novel hardware are
available that are less onerous on the
software~\cite{10.1145/2830772.2830802,9138965,8834719,7446055}. The
choice of approach to crash-consistency is not dictated by MCAS and is
left up to the ADO plugin developer. In our use case implementation
described later in Section~\ref{sec:usingmcas}, we use a modified C++
Standard Templates Library to provide transparent undo logging
together with 64-bit atomic operations.
\section{Applying the MCAS-ADO approach to a CDP Index}
\label{sec:usingmcas}

Ransomware has rapidly emerged as the most visible cybersecurity risk
threatening business, government and private sectors. It is a class
of malware that allows the attacker to encrypt data in order to then
extort the victim into paying a ransom, normally through
crypto-currency, for its decryption and restoration.

A key strategy in combating ransomware is to ensure that data is
copied into a secure storage tier, thereby allowing data to be
recovered in the event of attack. Continuous Data Protection (CDP)
refers to transparently making a backup of data in real-time and thus
maintaining an immutable history of the data which can be used for
restoration while minimizing data loss~\cite{SNIA17cdp}.

\begin{figure}
\centering
\includegraphics[width=\linewidth]{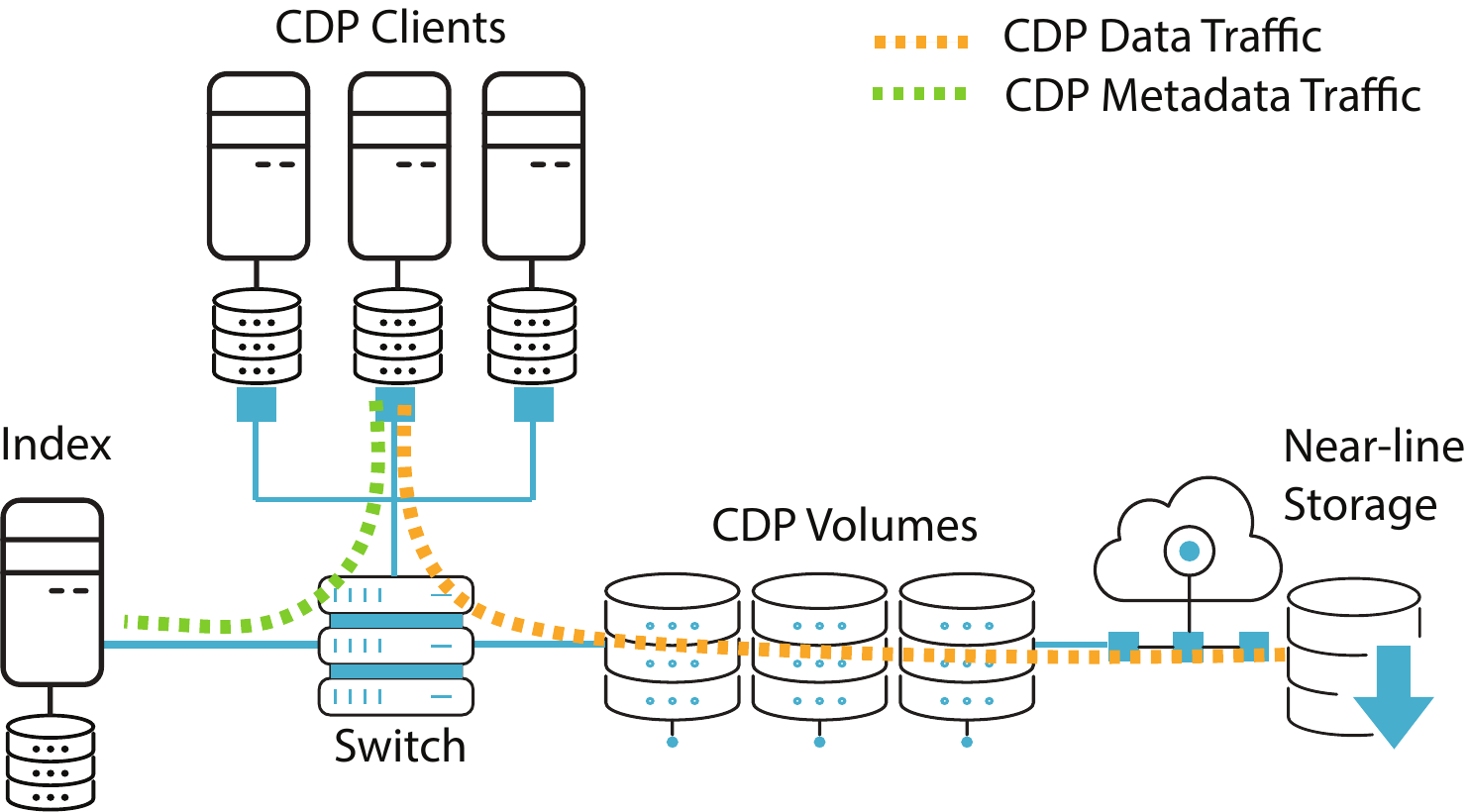}
\caption{Basic CDP Architecture}
\label{fig:cdparch}
\end{figure}

The focus use case, for our evaluation of the MCAS-ADO approach, is
achieving higher performance and scalability for a CDP index
implementation.  The broader CDP solution (outside the scope of
this paper) is to combine a distributed block storage system with a
high-performance, durable (replicated) index that maintains a mapping
between \textit{virtual blocks} and \textit{managed blocks} over
time (see Figure~\ref{fig:cdparch}).  Virtual blocks are those that
the application ``sees'', albeit through a filesystem in most
cases. As with locally attached storage, the virtual space only
represents the latest version. 

Managed blocks collectively represent versions over time. When the
application state needs to be rolled-back to a prior version during
recovery, the virtual space for that point-in-time is reconstructed
from the managed blocks.

In terms of normal behavior outside of an attack, the index must be
updated for each write in the system. For example, 10,000
containerized applications issuing 200 IOPS each (typical for a DB
application) requires 2M updates/second on the index.

The low-latency and guaranteed synchronous persistency provided by
MCAS enables the index to be kept consistent with the storage system
at fine granularity, while avoiding loss of any data in the event of
power failure or reset event.  Ultimately, this reduces complexity and
recovery time by avoiding the need to reconstruct the complete or
partial index in memory when the system starts up.  Furthermore, the
high update performance and smaller memory footprint, which is
a key element of our design, makes scaling to thousands
of applications possible.

\paragraph{The CDP Index}
\label{sec:cdp}

As previously discussed, the basic premise is to map a set of virtual
blocks to managed blocks over time. 

\begin{figure}[h]
\centering
\includegraphics[width=\linewidth]{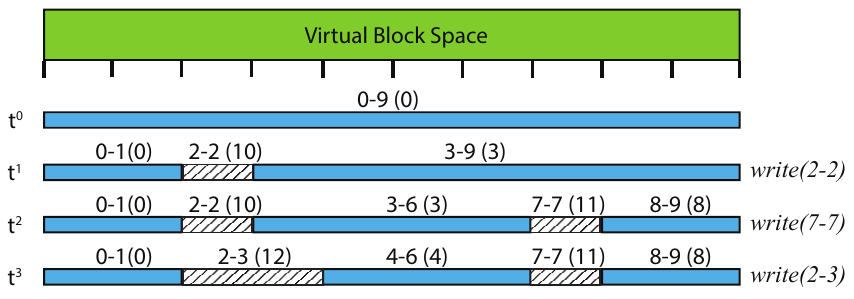}
\caption{Example Block Mapping over Time}
\label{fig:blkmap}
\end{figure}

Figure~\ref{fig:blkmap}
illustrates a simple example where three writes create new mappings to
the managed block space. The numbers in parenthesis designate the
offset in the managed space for the given mapped virtual range. In
the example, each new write is appended to the managed space. Mapped
ranges can represent one or more blocks and they do not overlap.

\begin{figure}
\centering
\includegraphics[width=1.0\linewidth]{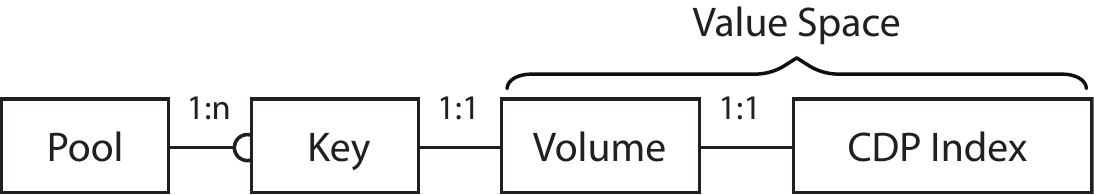}
\caption{Data Schema}
\label{fig:schema}
\end{figure}

\begin{figure}
\centering
\includegraphics[width=1.0\linewidth]{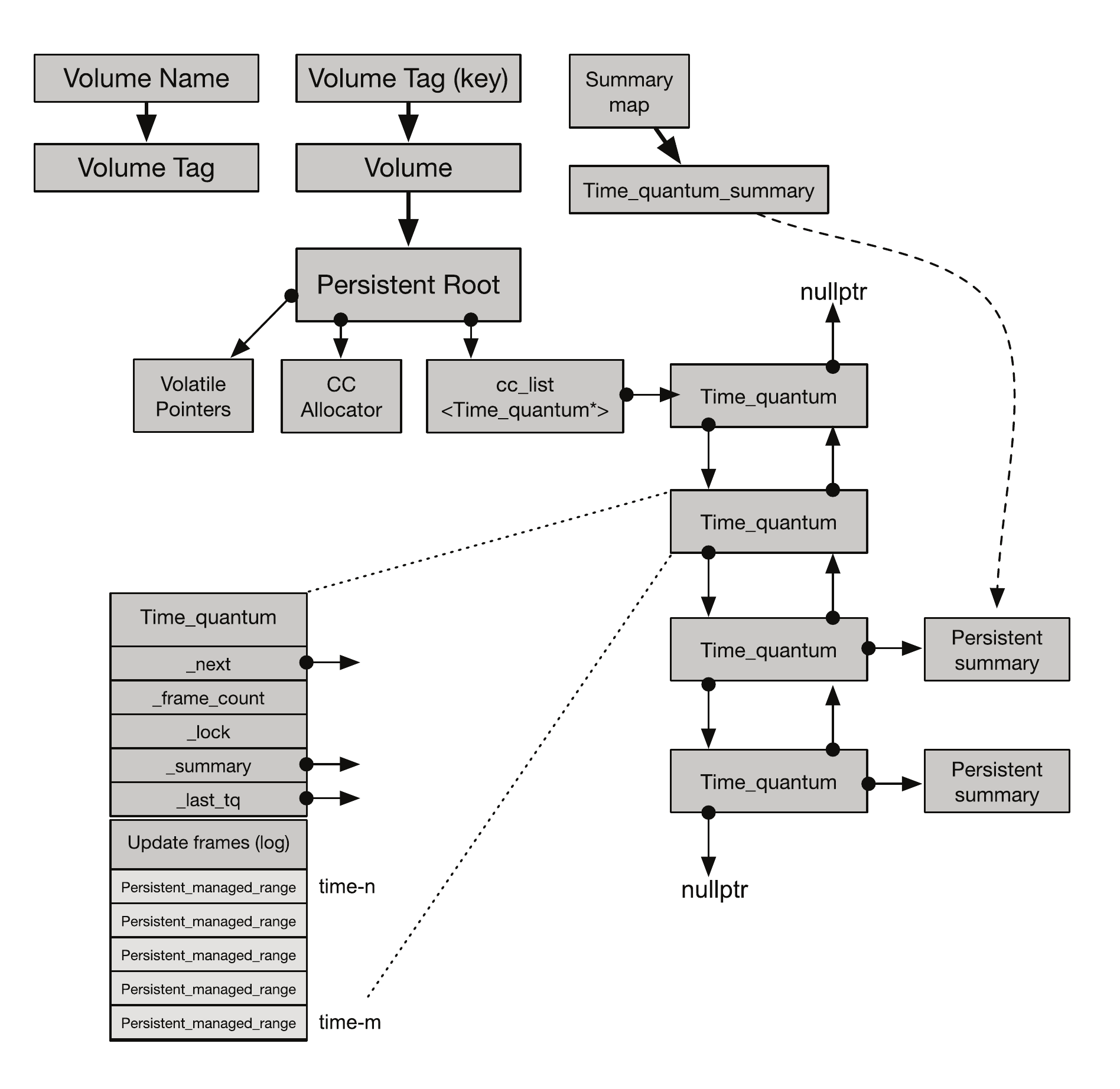}
\caption{CDP ADO Data Structure}
\label{fig:cdpdata}
\end{figure}

In order to scale to large containerized environments supporting
thousands of persistent volumes, we implement the CDP use case using a
2-level index organization.  At the top-level, each entry in a
hash table represents a key-value pair. The key is a Volume Tag (a
shortened version of a longer name) which uniquely identifies an
application volume (see Figure~\ref{fig:schema}). The value stores the
handle to a persistent data structure representing the CDP Index for
that volume. The CDP Index, described in Figure~\ref{fig:cdpdata},
records all mappings from virtual to managed blocks over time. Each
time a block is modified a new entry is created in the CDP Index.

In order to support millions of updates per second, update operations
must exhibit low latency and must be durable; lazy persistency or lazy
replication lead to potential data loss and more recovery complexity.
To support this performance requirement we leverage the flexibility
of the ADO plugin mechanism, which allows arbitrary pointer-based data
structures to be instantiated and modified.

From a logical point of view, the index maintains an up-to-date block
map for every point-in-time. However, to reduce the overall memory
footprint and compute demand, our design batches updates and performs
materialization of the map only on completion of the batch.  Each
batch referred to as a \code{Time\_quantum} contains updates that occur
during a time range.

Each write adds a 
\code{Persistent\_managed\_range} 64-byte
record\footnote{Being equal to the size of a cache line helps minimize
 read-write amplification, although some amplification still happens internally
 to the NVDIMM.} to the current \code{Time\_quantum}. 

Each record write is crash-consistent and made transactional 
through a combination of undo-logging and atomic operations. 

\subsection{Lazy Summarization and Point-in-time Query}

The memory size for each quantum is configurable. However, between
4MiB (64K records) and 16MiB (262K records) is typical. On reaching
full capacity, each quantum is ``handed over'' to a secondary thread
in the ADO plugin that performs \textit{lazy summarization}. The
summarization process builds the mapping state for the last point in
time in the quantum (i.e. the time point of last record). This is
achieved by taking the prior quantum's summary (held in DRAM) and
merging in all of the writes from the quantum in time order. The merge
process is performed in DRAM and on completion
transactionally copied and made durable in persistent memory.

Memory for the persistent data structure elements is allocated by a
local heap in the ADO plugin. The heap is populated with coarse-grained
regions of memory (chunks) to manage, that are allocated from the
memory pool via a call-back to the shard process. The exact size of
the chunks is configurable (64MiB is typical).

The fundamental purpose of the index is to provide a block mapping for
a specific instance in time, \(t\). After an attack event, the CDP
recovery process iteratively explores mappings for multiple points in
time. Mapping requests may be for the full block-range or some
narrower sub-range of the volume. The algorithm used to iteratively
determine an effective recovery point is outside the scope of this
paper.

Retrieval of a mapping for time \(t\) is as follows:

\begin{enumerate}
 \item Starting from the most recent, scan each element
  in the quantum list until a \code{Time\_quantum} \(Q_t\) is identified that has a timestamp less
  than or equal to \(t\).
 \item If the identified quantum does not have a summary (recall lazy
  summarization), continue searching backwards until a quantum with
  complete summary \(Q_s\) is identified (\(time(Q_s) \leq time(Q_t)\)).
 \item Starting with summary \(Q_s\) merge in all updates
   (\code{Persistent\_manage\_range} records), in time order, into a
   result \(R\). Continue merging updates for quantums up to \(Q_t\).
 \item Merge all records \(r \in Q_t\) while \(time(r) \leq t\) into result \(R\).
\end{enumerate}

Because lazy summarization is able to keep up with the incoming
updates, \(Q_t\) and \(Q_s\) are in most cases the same quantum. If the
query is for a sub-range, then updates are filtered by that range
(i.e. included if they overlap).

\subsection{Aging Out Data}

To guard against delayed attack detection, mapping data is retained
for a period of time. As the data's age extends beyond a pre-defined
threshold (e.g. 7 days) the mapping data is copied to a
slower tier and erased in MCAS. Our
design uses a secondary ADO plugin thread to ``trim'' the \code{Time\_quantum}
list periodically. This is done by scanning the crash-consistent
\code{Time\_quantum} list (\code{cc\_list}) and then removing the tail element
should its timestamp indicate that it is sufficiently old. After removal from the
list, the memory for the quantum and its associated summary are
released back to the ADO-local heap.

\subsection{Data Replication}

To protect against failures within a data center, our CDP use case also
requires that multiple replicas of the mapping data exist. 

The CDP plugin uses a client-side replication strategy.  Achieving
consistency requires that write-updates for a given volume are sent
from a single client, which holds true in our broader design. The client
does not send the next update until it has received an acknowledgement
for the prior update from all nodes.  Although details of failure
handling, reconciliation and recovery are outside the scope of this
paper, we include in the evaluation the write throughput using two-
and three-way replication demonstrating the ability of MCAS to sustain
high-performance even with synchronous replication.

\section{Evaluation}
\label{sec:eval}

In this section, we present our experimental results for the CDP use
case. We examine performance of both non-replicated and replicated
update (write) workload as well as sporadic point-in-time query
latencies (read).  In the current design, the client maintains a local
cache with the latest block mappings. A majority of the reads are
directly served out of the local cache.  Hence a 100\% query workload
is not a useful measure for this use case.  On the other hand, every
write operation results in a write operation to the MCAS backend to
update the index. As a result, write throughput
(Section~\ref{sec:rwt}) is a key measure of performance.

In order to evaluate the benefit of near-data compute, we compare
the CDP ADO performance with a ``thick'' client alternative that uses MCAS
as a basic key-value store and performs CDP quantum summarization and 
query handling on the client node.

\begin{figure}[]
\centering
\includegraphics[width=1.0\linewidth]{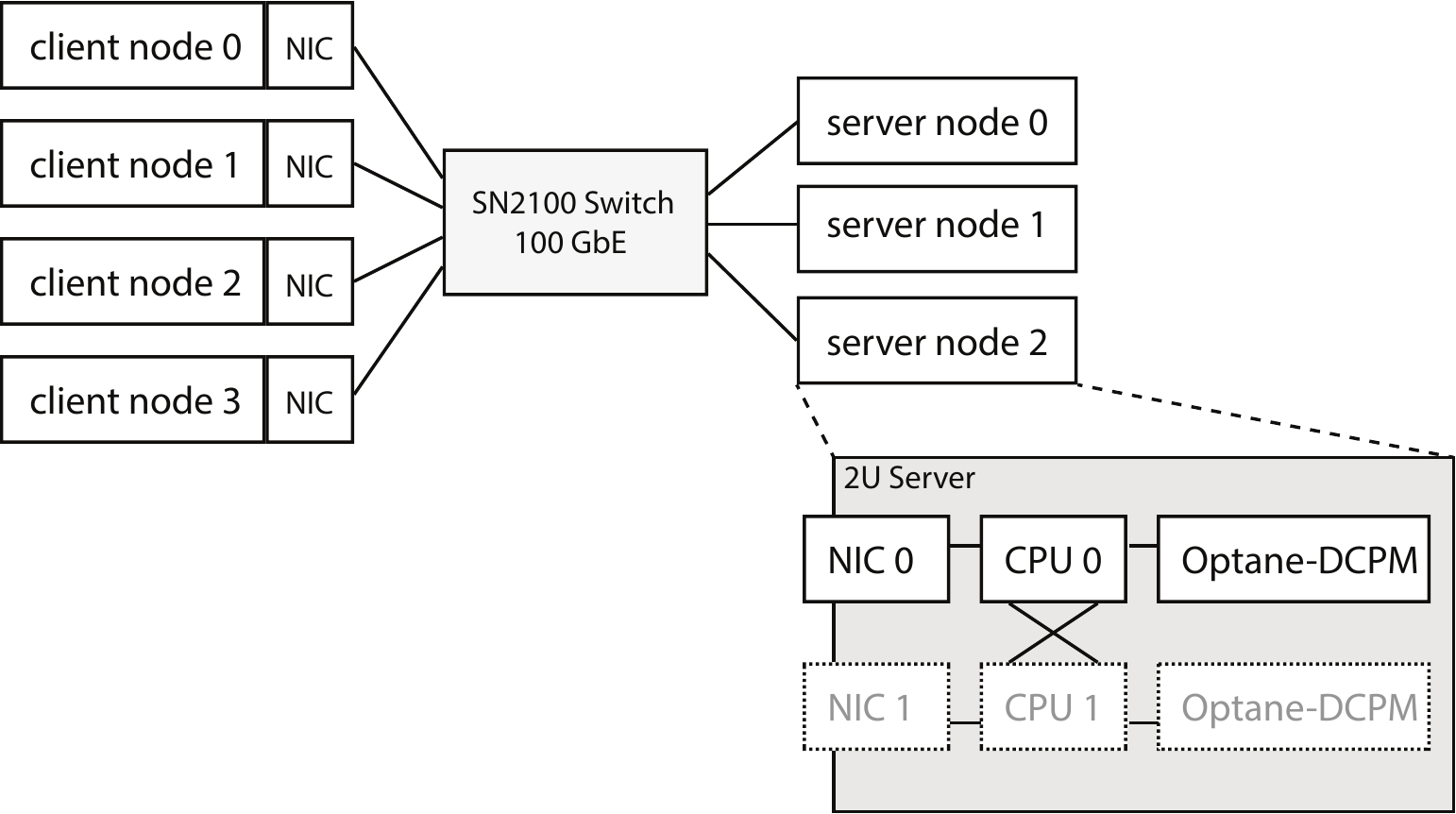}
\caption{Experimental Network Topology}
\label{fig:net_top}
\end{figure}

Table~\ref{tab:spec} details our experimental hardware and software
configurations. Figure~\ref{fig:net_top} shows the network topology.
Note that the scaling experiments use a single processor, with
NUMA-local memory, and a single NIC.

\begin{table}[t]
\begin{centering}
\begin{tabularx}{1.0\linewidth}{
>{\setlength{\hsize}{.2\hsize}\raggedright\footnotesize}X
>{\setlength{\hsize}{.8\hsize}\raggedright\arraybackslash\footnotesize}X }
 \hline
 \textbf{Component} & \textbf{Description} \\
 \hline
 Processor & Intel Xeon Gold 5128 (Cascade Lake) 16-core 2.30GHz \\
 Cache & L1 (32KiB), L2 (1MiB), L3 (22MiB) \\
 DRAM & PC2400 DDR4 16GiB 12x DIMMs (192GiB) \\
 NVDIMM & \optane 128GB 12x DIMMs (1.5TB) \\
 RDMA NIC & Mellanox ConnectX-5 (100GbE) \\
 OS & Linux Fedora 27 with Kernel 4.18.19 x86\_64 \\
 Compiler & GCC 7.3.1 \\
 NIC S/W & Mellanox OFED 4.5-1.0.1 \\ 
 \hline
\end{tabularx}
\caption{Server system specification}
\label{tab:spec}
\end{centering}
\end{table}

Each pinned shard thread has a 1:1 association with a pinned ADO
primary thread.  A fraction of threads are kept floating to serve as
ADO secondary threads.  These are used to perform background
operations such as summarization and age-out.  A fully populated
single-socket that supports 32 cores (using Hyper-Threading) will
support up to 12 shards, 
12 primary ADO
threads and 8 secondary threads.  A two-socket server can support
twice this.

\subsection{Index Write Performance}
\label{sec:rwt}

This experiment measures the maximum write throughput and scaling
across an increasing number of shards. Clients generate random mapping
updates for volumes spanning 1M blocks with a random span of 1-100
blocks. Update timestamping uses actual time (nanosecond epoch).
Quantums are 64K records and each is aged out after a maximum of 10
newer quantums exist. Each client uses 1, 2 or 6 threads
executing updates to different volumes. Note that 6 threads is
sufficient to saturate performance on a single shard. The throughput
mean is taken over a 10 minute period.

\begin{figure}[h!]
\centering
\includegraphics[width=1.0\linewidth]{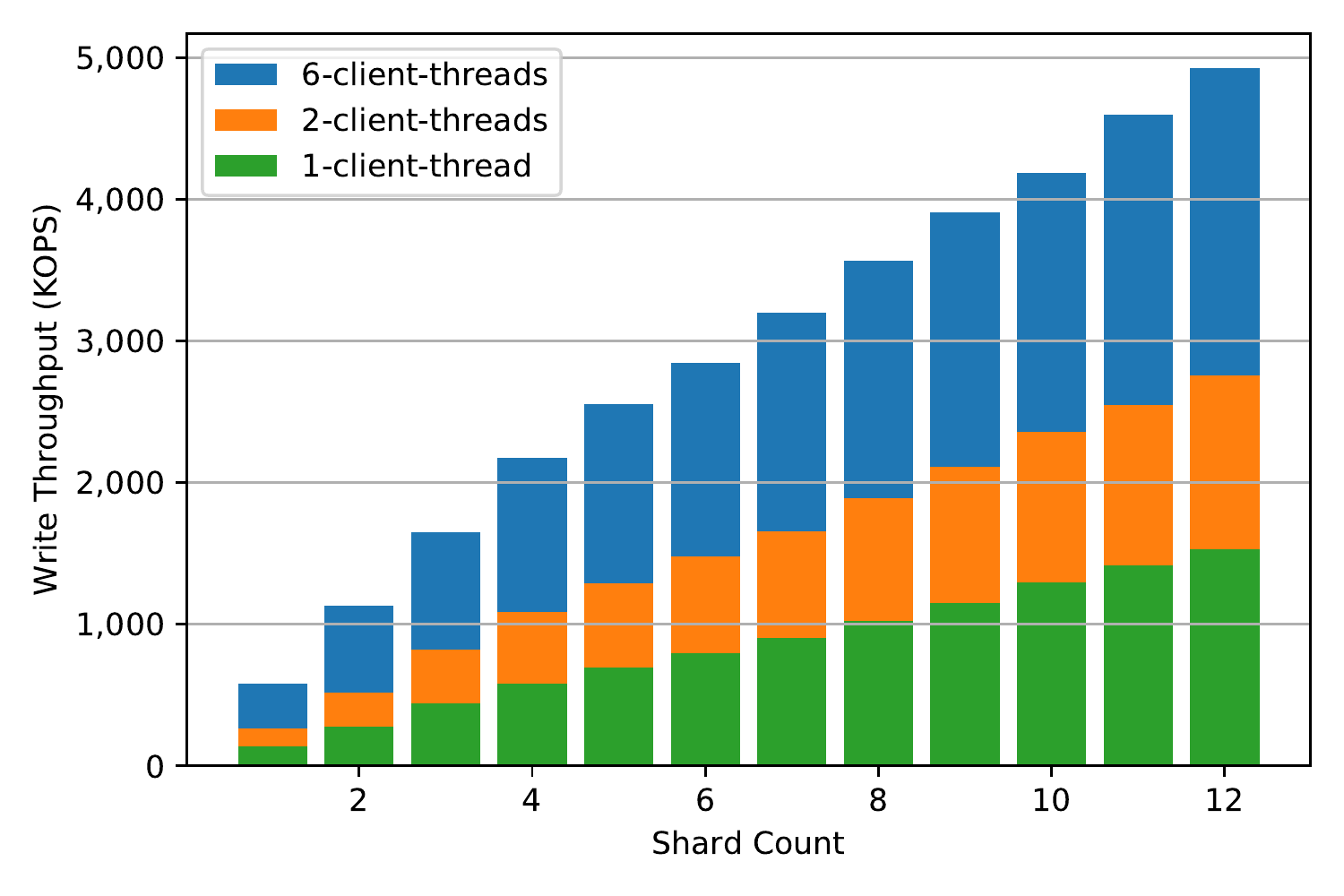}
\caption{Throughput Scaling without Replication}
\label{fig:perf0-1}
\end{figure}

For the update workload, Figure~\ref{fig:perf0-1} shows throughput scaling
of up to 4.92M updates/second, while a single client thread can achieve
140K updates/second with a mean round-trip time of 7.14\(\mu sec\).

\paragraph{Client Observed Update Latency}

\begin{figure}[h!]
\centering
\includegraphics[width=1.0\linewidth]{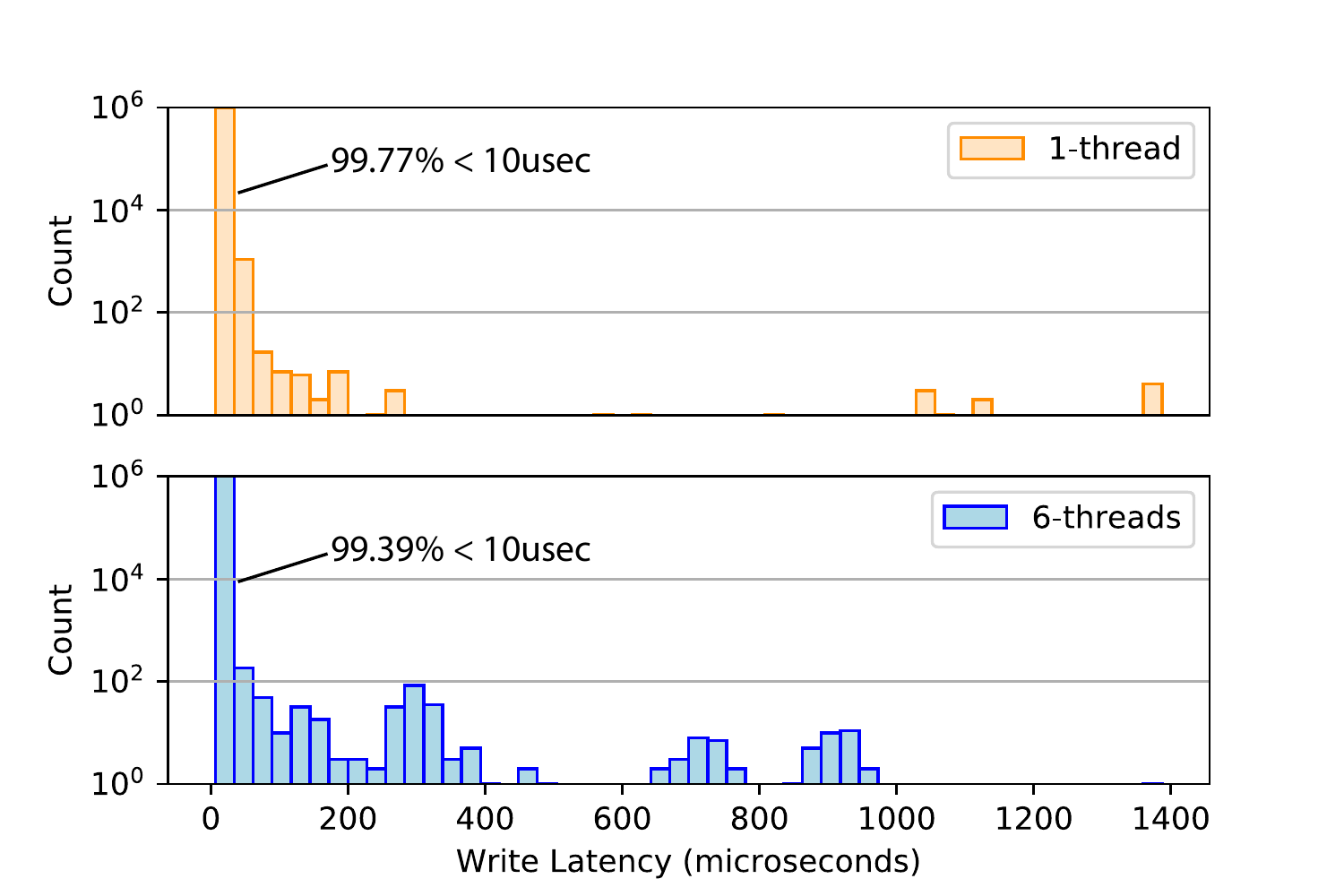}
\caption{Round-Trip Latency Distribution}
\label{fig:write-lat}
\end{figure}

Figure~\ref{fig:write-lat} shows latency distributions for a single
shard operating with 1 and 6 client-side threads. A logarithmic scale
is used for the Y-axis and outliers have not been removed. The data
shows that 99.77\% of the latency for the 1-threaded client is less
than 10\(\mu sec\), while 99.39\% of writes for the 6-threaded client
are also less than 10\(\mu sec\). As would be expected, latency for the
6-thread scenario is increased as threads compete in a queued
fashion.

We believe that this tight long-tail latency cannot be achieved
with other traditional storage technologies such as NVMe SSD.

\begin{figure}[h!]
\centering
\includegraphics[width=1.0\linewidth]{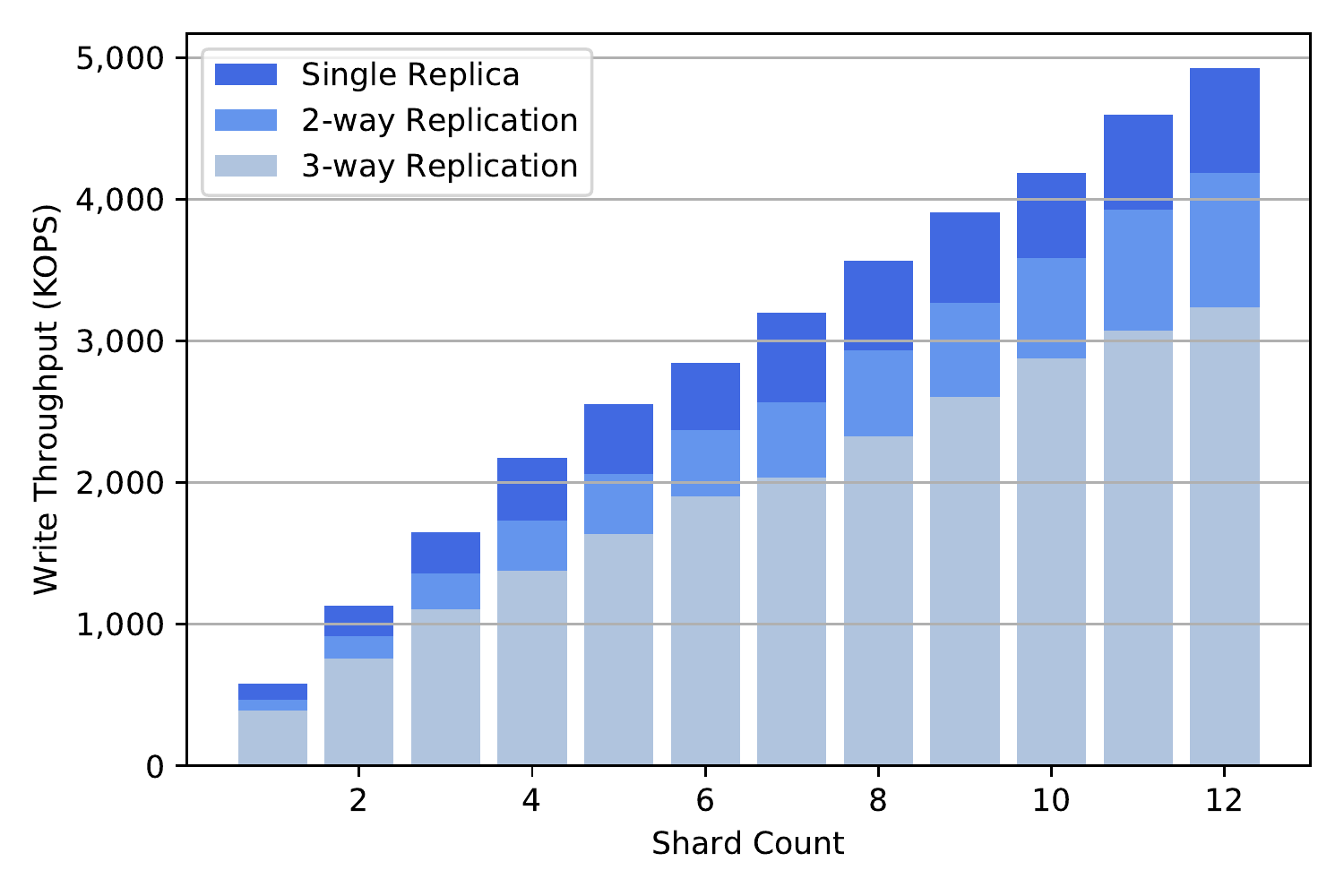}
\caption{Throughput Scaling with Replication}
\label{fig:cli-side-rep-scaling}
\end{figure}

\paragraph{Replication Scaling}
Using 6 threads per client, Figure~\ref{fig:cli-side-rep-scaling}
illustrates scaling 
for two- and three-way replication. For 12 shards, the system is able
to scale to 3.92M and 3.07M updates/second for two- and
three-way replication, respectively. This represents a degradation of
20\% and 37\%.

\subsection{Query Latency}
\label{sec:exp:querylatency}

We now explore the query latency characteristics of our solution. To
re-cap, a query consists of materializing a virtual-to-managed block
mapping at some point-in-time \(t\).  This is achieved by using the
time \(t\)'s prior quantum summary as a starting point and then
merging in updates up to time \(t\). Thus, the time taken to
materialize is dependent upon the size of the quantum and the position
of \(t\) within the quantum.

\begin{figure}[h!]
\centering
\includegraphics[width=1.0\linewidth]{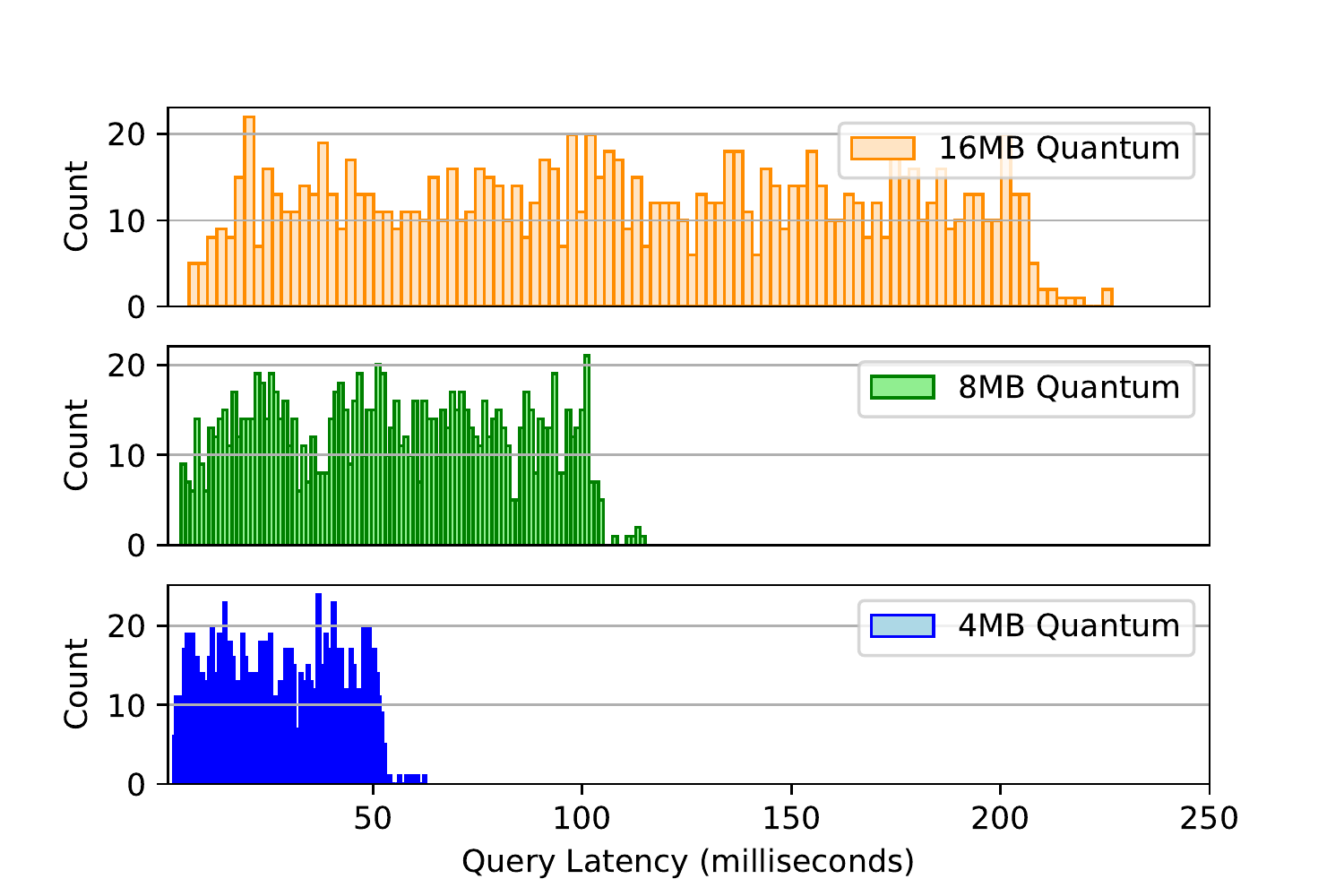}
\caption{Query Latency Distributions for Different Quantum Sizes}
\label{fig:query-lat}
\end{figure}

To illustrate this, Figure~\ref{fig:query-lat} shows latencies for
different quantum sizes: 4MiB (65K records), 8MiB (131K records) and
16MiB (262K records). Time points are randomly selected and the
mapping for a contiguous region of 100K blocks is retrieved. The data
shows that for each 4MiB increase in quantum size the worst case
latency is increased by approximately 50$ms$.

\paragraph{Query Under Load}

We also examine how queries impact on-going concurrent write streams.
In our use case, we expect that in most cases not all clients will
need to recover together; some clients will continue to function as
normal. To this end, we examine how throughput rate for a single
client (which because of the synchronous behavior correlates directly
with latency) is affected by another client making concurrent query
requests on the same shard.

\begin{figure}[h!]
\centering
\includegraphics[width=1.0\linewidth]{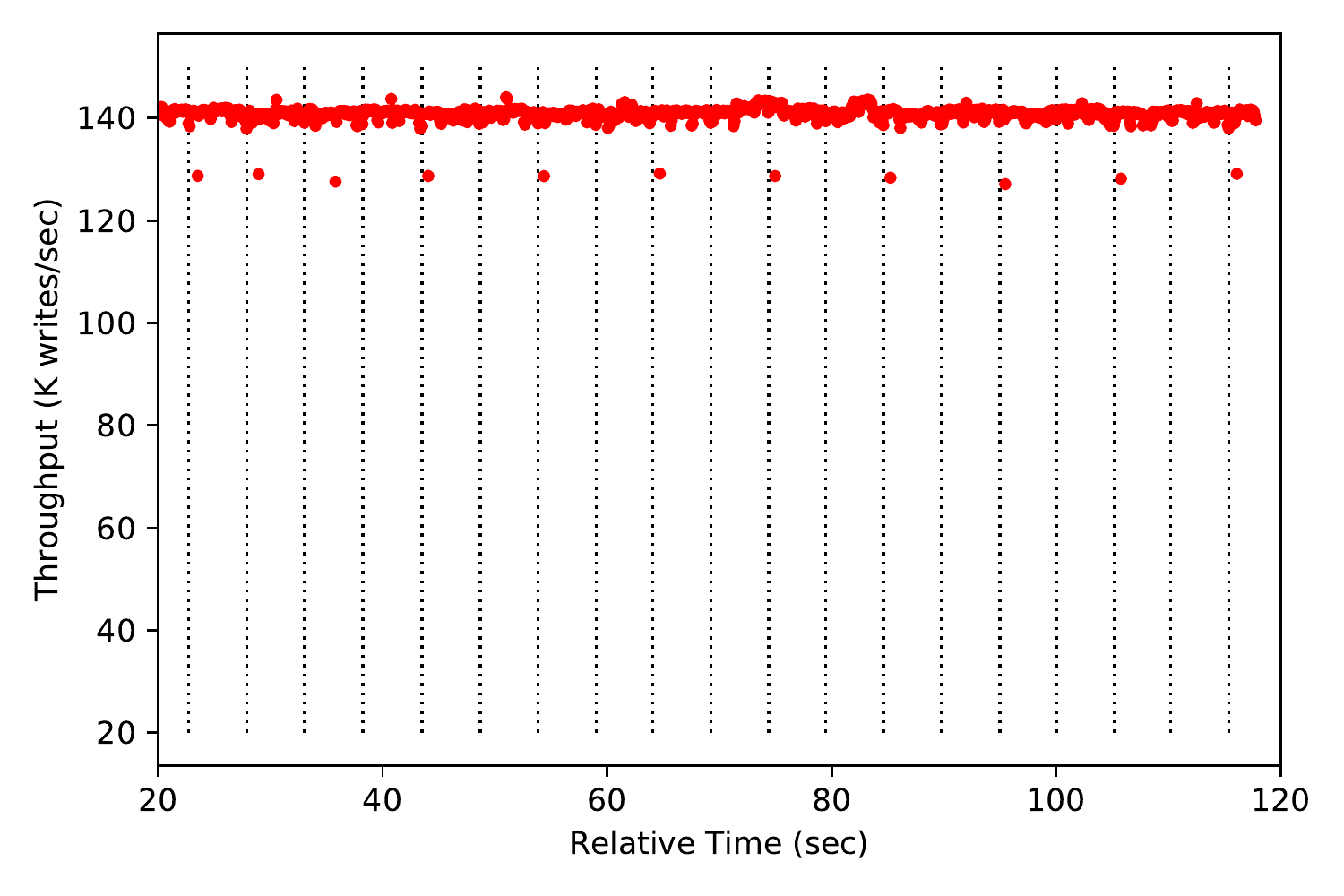}
\caption{Query Latency Under Load}
\label{fig:query-lat-load}
\end{figure}

Given writing client \(C_w\) and query client \(C_q\),
Figure~\ref{fig:query-lat-load} shows how the throughput of \(C_w\) is
impacted by periodic queries made by \(C_q\). The vertical dotted lines
mark the point-in-time at which query results are received by \(C_q\).
The red dots represent throughput samples at 12.5K updates intervals
by \(C_w\). The maximum throughput observed is 144K updates/second and
the minimum observed is 127K updates/second. Worst-case degradation
is thus 11.8\%.

\subsection{Comparison to Non-ADO Alternative Architecture}

In order to better validate and quantify the performance improvement
resulting from the MCAS ADO architecture, we compare the ADO-based solution
with a ``plain'' key-value solution that performs the same CDP
index operations but on the client side. To achieve a fair comparison,
we use the basic MCAS key-value store without the ADO
feature~\cite{waddington2020ispass}.

\begin{figure}
\centering
\includegraphics[width=1.0\linewidth]{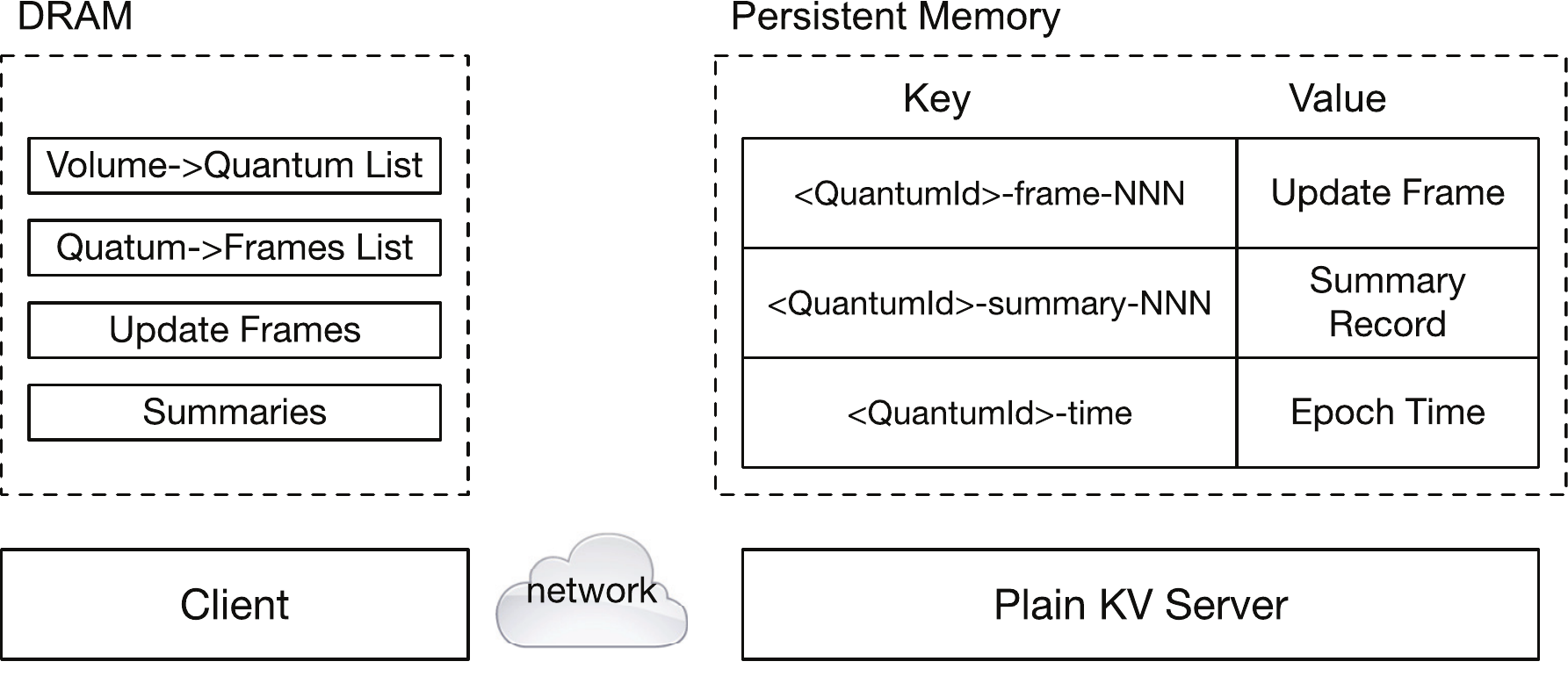}
\caption{Alternative Deployment with Plain Key-Value Store}
\label{fig:plainkv}
\end{figure}

Figure~\ref{fig:plainkv} shows the Plain-KV deployment. The CDP data
structures are broken down into individual key-value pairs. The primary
key is the \code{QuantumId}, which is derived from the Volume Name and
a unique numerical identifier.

Local copies of all data are maintained in the client's local
volatile memory (DRAM) and hence in practice would be limited to the
available client memory. Copies of the update frames and any summaries
are copied in the remote store's persistent memory for durability. However,
queries on the data are made from local copies alone (except when
performing client-side rebuild). Data age-out is
realized as explicit key-value pair deletions.

\begin{figure}[ht]
\centering
\includegraphics[width=1.0\linewidth]{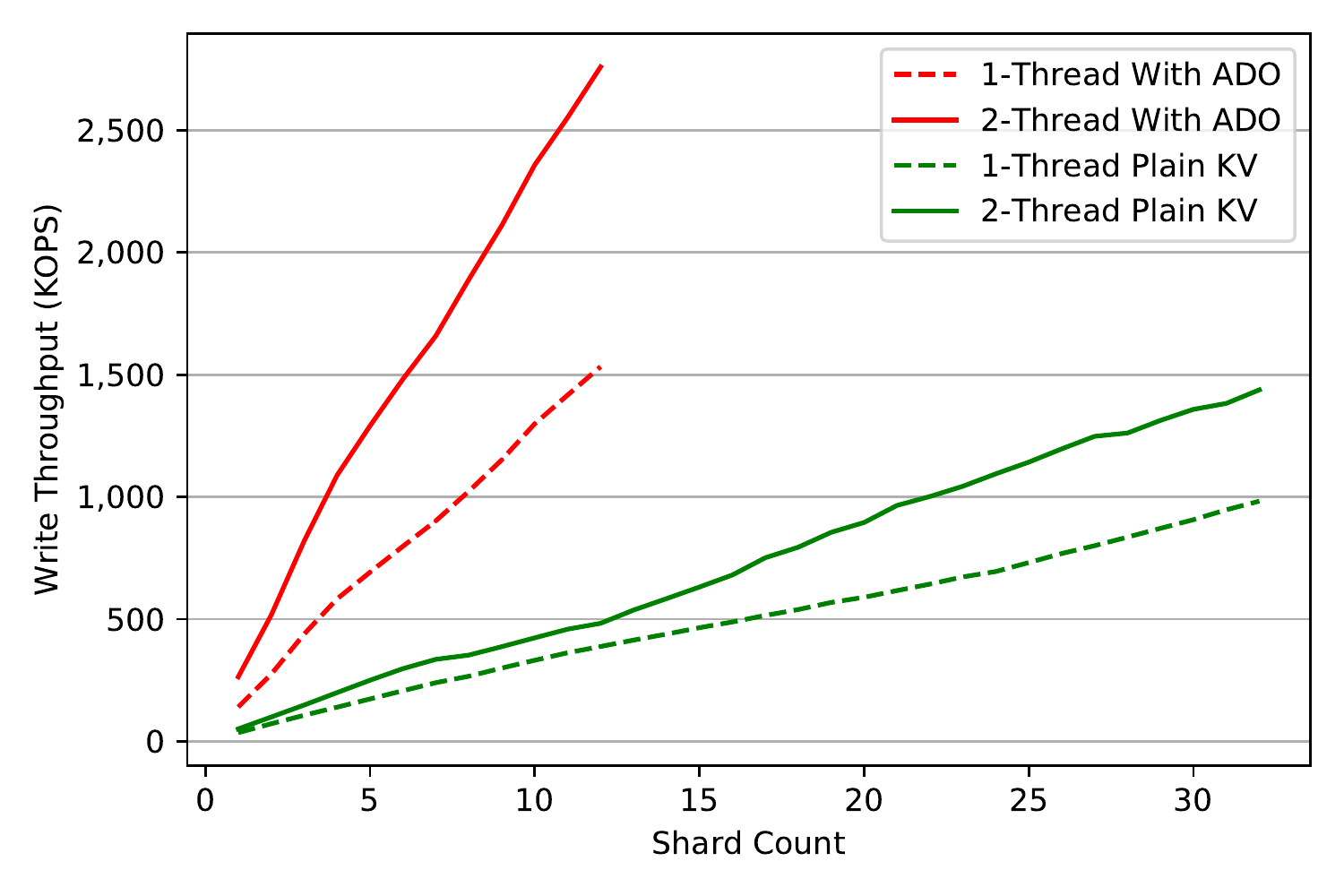}
\caption{Comparison of Throughput for ADO vs. Plain-KV Deployments}
\label{fig:plainkv-throughput}
\end{figure}

\paragraph{Throughput}

Figure~\ref{fig:plainkv-throughput} shows a comparison of throughput
under a 100\% write/update workload.

The Plain-KV deployment does not need ADO threads and therefore is
able to scale shards further to a maximum of 32 (at one thread per
shard). The figure shows 1 and 2 client threads; we do not have
sufficient test client machines available to provide data for 6
threads/client for Plain-KV. The data shows that the ADO-based
deployment, with the same total number of threads, can sustain 43\%
and 30\% higher throughput than Plain-KV for 2 threads/client and 1
thread/client respectively.

\begin{figure}
\centering
\includegraphics[width=1.0\linewidth]{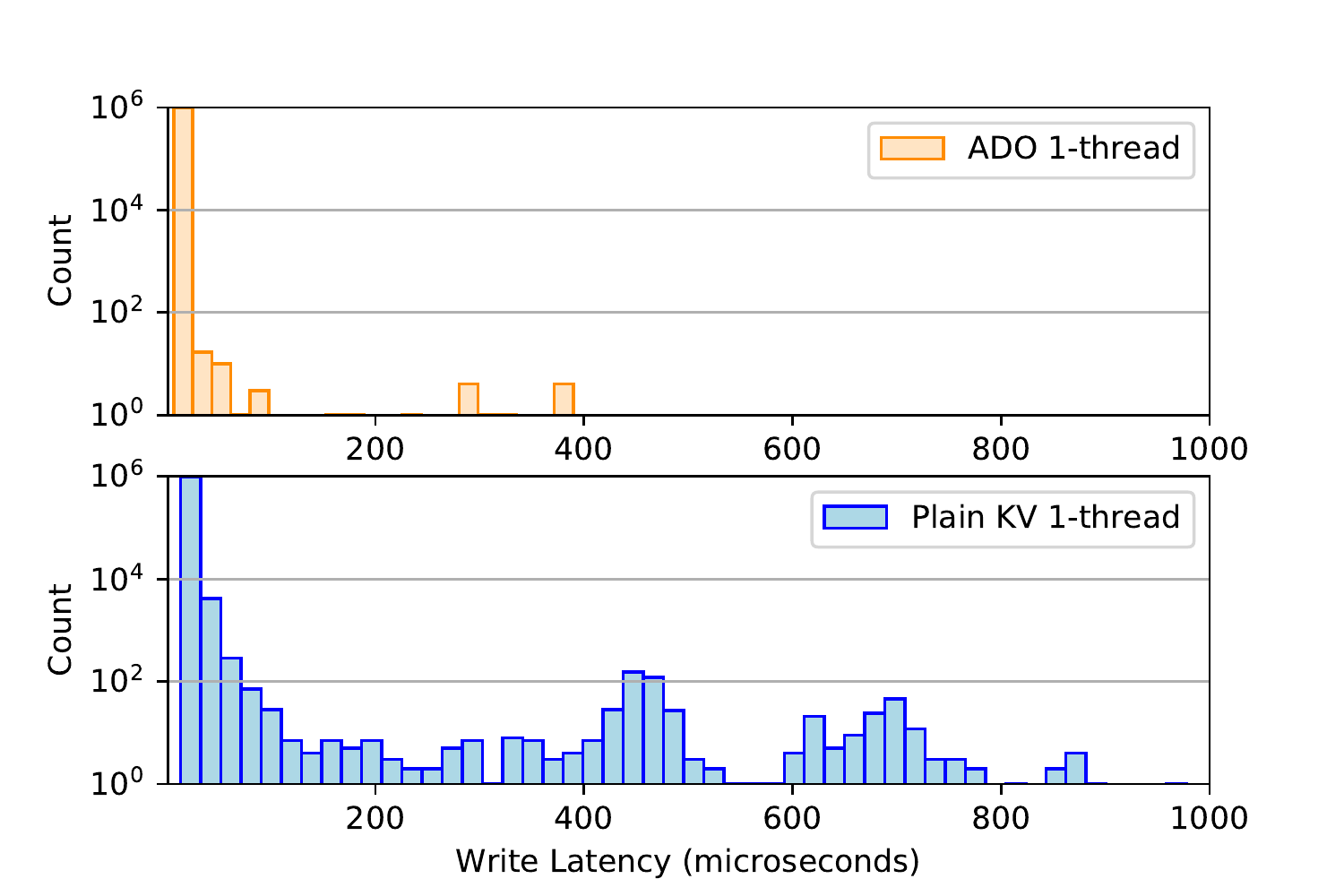}
\caption{Comparison of Write Latency for ADO vs. Plain-KV Deployments}
\label{fig:plainkv-latency}
\end{figure}

Figure~\ref{fig:plainkv-latency} compares the write latency
distribution for the two deployments. For the ADO deployment, 99.97\%
are less than 20\(\mu sec\) and for Plain-KV, 94.27\% are less than
20\(\mu sec\). The mean latencies are 6.7\(\mu sec\) and 16.6\(\mu sec\)
for the ADO and Plain-KV deployments respectively. The increased latency
observed by Plain-KV is due to the fact that in addition to remote
write updates, the single client thread is performing local data
structure updates, remote summaries updates and erases for aging-out
of the data (all on the same shard).

\paragraph{Query Latency}

\begin{figure}
\centering
\includegraphics[width=1.0\linewidth]{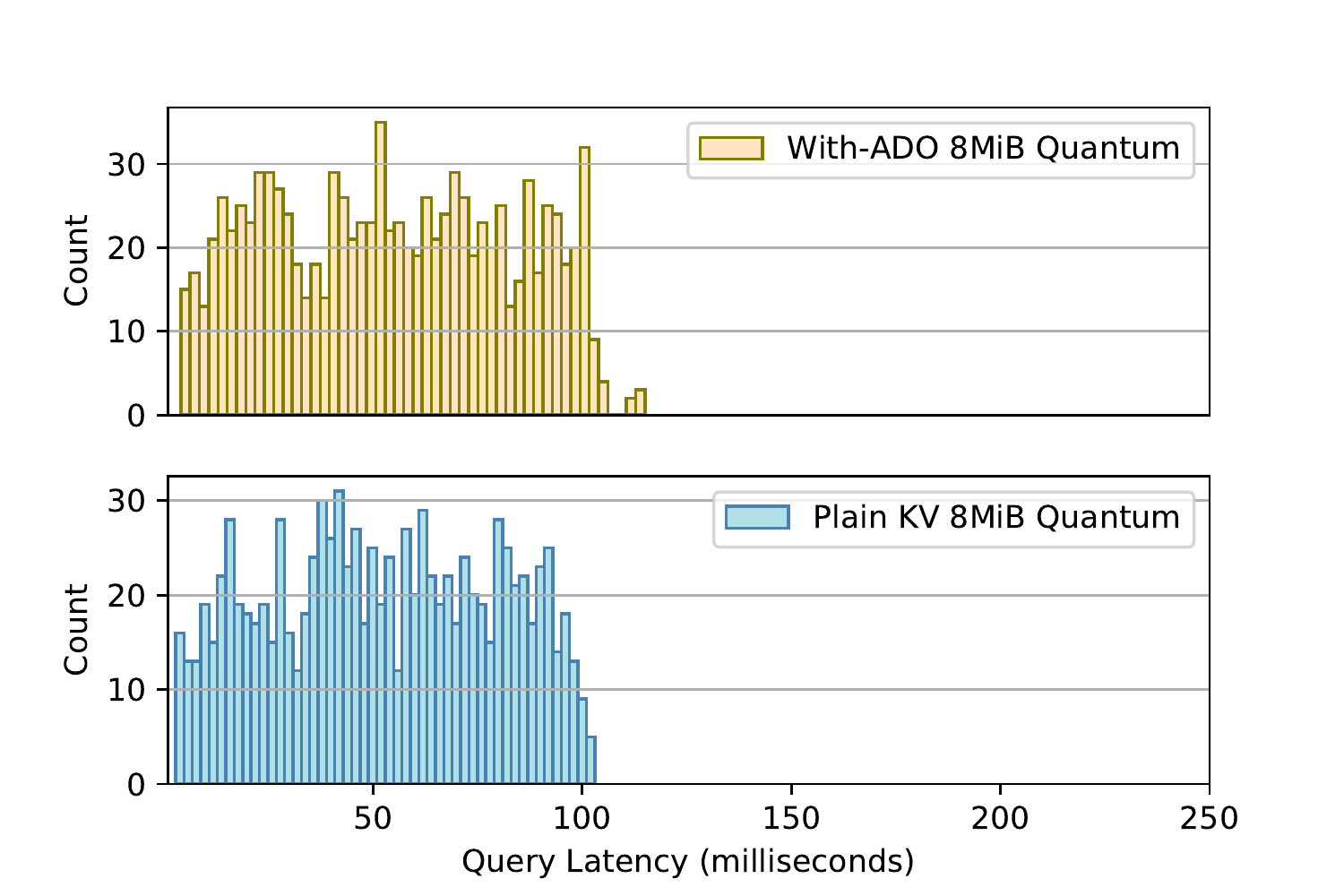}
\caption{Comparison of Query Latency for ADO vs. Plain-KV Deployments}
\label{fig:plainkv-query}
\end{figure}

Next, we compare the query latency for the two deployments.
Figure~\ref{fig:plainkv-query} compares latencies for an 8MiB quantum.
For the ADO deployment latencies vary from 4.0\(ms\) to 115.0\(ms\) with a
mean of 54.4\(ms\). Slightly better, the Plain-KV deployment shows
latencies varying from 2.8\(ms\) to 103.0\(ms\), with a mean of 52.8\(ms\).
We consider this improvement of negligible value in practice.

\paragraph{Memory Footprint}

Finally, we compare the memory footprint for each deployment. Quantums
are 16MiB (262K records) and they are aged-out at a limit of 10. A
single shard with single client thread arrangement is used.

\begin{table}[h!]
\begin{centering}
\begin{tabularx}{1.0\linewidth}{
>{\setlength{\hsize}{.3\hsize}\raggedright\footnotesize}X
>{\setlength{\hsize}{.3\hsize}\raggedright\footnotesize}X
>{\setlength{\hsize}{.4\hsize}\raggedright\arraybackslash\footnotesize}X }
 \hline
 & \textbf{ADO} & \textbf{Plain-KV} \\
 \hline
 Client DRAM   & 912 MiB & 1747 MiB (\(\uparrow\) 91\%) \\
 Server DRAM   & 530 MiB & 651 MiB (\(\uparrow\) 23\%) \\
 Server Optane PM & 327 MiB & 655 MiB (\(\uparrow\) 100\%)\\
 \hline
 Total Memory Footprint & 1.72 GiB & 2.98 GiB (\(\uparrow\) 72\%)\\
 \hline
\end{tabularx}
\caption{Memory Footprint Comparison}
\label{tab:footprint}
\end{centering}
\end{table}

The data in Table~\ref{tab:footprint} provides a comparison
of client and server side memory footprints for the different schemes.
Note, the seemingly large client-side footprint is due to
the replay data, which is pre-loaded before execution of the benchmark (in both instances).

The ADO deployment is more memory efficient mainly because there is only
a single copy of the data. However, the \optane footprint for
the Plain-KV deployment is twice that of the ADO deployment. This is
caused by the metadata overhead incurred by saving updates as
individual entries in the key-value hash table - in the ADO deployment
the quantum is a contiguous array of records in memory.

\section{Related Work}
\label{sec:related}

In this section, we review related work on the topics of high
performance in-memory key-value stores, use of persistent memory as
the storage medium and the use of in-store operations. 

\paragraph{In-memory key-value stores}
\label{sec:related:inmem}

Stanford University's RAMCloud~\cite{10.1145/2806887} is a distributed
key-value store that exploits DRAM for high-performance storage
together with replication to provide durability. It focuses on
maintaining in-memory low-latency while also achieving scalability and
durability. Similarly,
FaRM~\cite{10.5555/2616448.2616486,10.1145/2815400.2815425},
HERD~\cite{10.1145/2619239.2626299} and
MICA~\cite{10.5555/2616448.2616488} further optimized latency by
leveraging high-performance RDMA for networked in-memory key-value
stores. All of these are designed for DRAM and have demonstrated
sub-millisecond latencies. However, they are not able to manage
crash-consistency and persistent data structures.

\paragraph{Persistent memory aware data stores}

Other work takes existing data stores and modifies the back-end to
support \optane, which means replacing read/write I/O instructions
with direct load/store memory instructions. Izraelevitz
et. al~\cite{izraelevitz2019basic} change the back-end of Redis,
RocksDB, and MongoDB to support \optane.

The general outcome from this work is a demonstration of better
performance when using persistent memory directly with DAX mode (as
opposed to block mode).

More recently, PMemKV is a new \optane-optimized key-value
data-store~\cite{PMemKV} developed by Intel. Its storage engine can be
replaced by various tree data structures provided by PMDK~\cite{PMDK}.
Izraelevitz et. al~\cite{izraelevitz2019basic} benchmarked different
storage engines and measured random read latency at around 1\(\mu sec\)
and write latency at around 4\(\mu sec\). Note, this benchmark provides
only local performance for a single thread; the system is not networked.

\paragraph{Remote memory paradigm}

AsymNVM~\cite{10.1145/3373376.3378511} focuses on providing RDMA-like
operations but with transactional semantics.  The key idea to remove
persistency bottleneck is the use of an operation log that reduces
stall time due to RDMA writes and enables efficient batching and
caching in front-end nodes.  AsymNVM is based on a remote memory
paradigm; there is no index or key space. Furthermore, AsymNVM
does not allow operations close to the memory.

Clover~\cite{Tsai2020DisaggregatingPM} offers an exploration of
passive disaggregated persistent memory. They do provide space
management but their focus is primarily a remote memory paradigm in
that compute is performed on the client side and memory is attached
remotely.

StRoM (Smart Remote Memory)~\cite{10.1145/3342195.3387519} investigates
the use of FPGA-based computation in the smart NIC card to perform operations
on remote memory.  Limited by the FPGA, they provide only fundamental
operations on the data (e.g., pointer-chasing, filtering, aggregation).

\paragraph{In-store operations}

Prior work in ``shipping'' operations to a data store focuses on
using volatile memory combined with conventional storage. To our
knowledge, MCAS is the first work to support in-store operations in
persistent, byte addressable memory.

Stored procedures are widely adopted in the database community. They
are typically in the form of programs defined in a programming
language that are translated (as opposed to natively compiled) at
runtime~\cite{TranscatSQL, PLSQL}. To improve performance, stored
procedures can be translated prior to execution into compact and more
efficient machine code~\cite{Hekaton}.
H-Store~\cite{10.14778/1454159.1454211},
VoltDB~\cite{Stonebraker2013TheVM} and
Hazelcast~\cite{10.1145/3030207.3053671} are examples of databases
that use Java-based stored procedures.

Redis~\cite{10.5555/2505464} pioneered early work on supporting
user-defined extensions in key-value stores. Redis allows customized
modules with built-in data structures. The modules are able to
manipulate data structures directly in the store. However, Redis is
limited to built-in data structures in contrast to MCAS's ability
to support any native pointer-based data structure.

Splinter~\cite{KulkarniMNZRS18} is another in-memory key-value store
that supports natively executable extensions. It supports any
customized operations on arbitrary data structures written in
Rust~\cite{rust}. However, unlike MCAS, user-defined extensions in
Splinter cannot directly operate on value data structures (it does not
use memory mapping). Instead, it copies data to commit changes. Splinter
also uses DRAM and therefore does not need to deal with data
persistency and crash-consistency.

\section{Conclusion and Future Work}
\label{sec:con}

In this paper we have outlined an evolution of the traditional
key-value store paradigm with near-data compute in the form of Active
Data Objects.  To our knowledge, this is a first-of-its-kind
technology that combines RDMA networking, persistent memory and
crash-consistent in-store programmability. We have evaluated our
approach in the context of a real-world use case based on indexing for
Continuous Data Protection (CDP).  We show that the ADO-based approach
results in up to 43\% higher throughput performance and 72\% less
memory footprint compared to the Plain-KV approach.  In terms of raw
performance numbers, with three-way replication, our solution can over
~6M updates/sec in a three node cluster of 2-socket servers.

The low read-write amplification and low latency enabled by \optane, in
conjunction with RDMA, allows the system to support a synchronous
``memory like'' behavior that simplifies design and consistency
models. With three replicas (immediately consistent), MCAS can
potentially support thousands of persistent volumes with an aggregate
IOPS throughput of over 6M IOPS in a 2-socket server.

A key design element of MCAS is its Active Data Object (ADO) plugin
mechanism. This feature allows the system to extend the key-value
paradigm with arbitrary in-place operations that can be safely
executed (using process isolation) in the MCAS system. In the current
prototype, users implementing ADO operations use either hand-crafted code
or STL instrumented with undo log capabilities. We believe that a key
hurdle for the adoption of persistent memory is making it easier to
develop crash-consistent code. 

From the perspective of furthering
our solution, we plan to explore the use of new languages, compilers and
libraries as well as hardware support for transparent undo
logging~\cite{9138965,7446055}. Extending these concepts to allow for
secure (e.g. enclave-like) execution in the storage system is also a
topic of further work.

\section*{Availability}

MCAS is an open source project maintained at
\code{https://github.com/IBM/mcas/}.  Additional information about the
project can be gained from \code{https://github.com/IBM/mcas/}.

\bibliographystyle{plain}
\bibliography{references}

\end{document}